\documentclass[11pt]{article}
\usepackage{scrextend}
\usepackage{style}
\usepackage{times}
\usepackage{url}
\usepackage{latexsym}
\usepackage{graphicx}
\usepackage{caption}
\usepackage{enumitem}
\usepackage{subfig}
\usepackage{appendix}
\usepackage{amssymb}
\usepackage{amsmath}
\usepackage{svg}
\usepackage{multirow}
\usepackage{float}
\usepackage{tablefootnote}
\setsvg{inkscape = inkscape -z -D}
\setsvg{svgpath = figs/}
\eaclfinalcopy

\title{Fine-Grained Named Entity Recognition using ELMo and Wikidata}

\author{
  {\bf Cihan Dogan, Aimore Dutra, Adam Gara, Alfredo Gemma}, \\ 
  {\bf Lei Shi, Michael Sigamani, Ella Walters }\\
  Constellation AI \\
  7 Carlisle Street \\ 
  London, W1D 3BW, United Kingdom \\
  {\tt michaelsigamani@constellation.ai} \\
}
\date{}

\begin{document}
\maketitle
\begin{abstract}

Fine-grained Named Entity Recognition is a task whereby we detect and classify entity mentions to a large set of types.
These types can span diverse domains such as finance, healthcare, and politics. We observe that when the type set spans several domains the accuracy of the entity detection becomes a limitation for supervised learning models.
The primary reason being the lack of datasets where entity boundaries are properly annotated,
whilst covering a large spectrum of entity types.
Furthermore, many named entity systems suffer when considering the categorization of fine grained entity types.
Our work attempts to address these issues, in part, by combining state-of-the-art deep learning models (ELMo) with an expansive knowledge base (Wikidata). Using our framework, we cross-validate our model on the 112 fine-grained entity types based on the hierarchy given from the Wiki\textsc{(gold)} dataset.

\end{abstract}

\section{Introduction}

Named entity recognition (NER)~\cite{collins1999unsupervised,tjong2003introduction,ratinov2009design,manning2014stanford}
is the process by which we identify text spans which mention named entities, 
and to classify them into predefined categories such as
\textit{person}, \textit{location}, \textit{organization} etc.
NER serves as the basis for a variety of natural language processing (NLP) 
applications such as relation extraction~\cite{mintz2009distant}, machine translation~\cite{koehn2007moses},
question answering~\cite{lin2012no} and knowledge base construction~\cite{dong2014knowledge}.
Although early NER systems have been successful in producing adequate recognition accuracy,
they often require significant human effort in carefully designing rules or features.

In recent years, deep learning methods been employed in NER systems, 
yielding state-of-the-art performance. However, 
the number of types detected are still not sufficient for certain domain-specific applications.
For relation extraction, identifying fine-grained types has been shown 
to significantly increase the performance of the extractor~\cite{ling2012fine,koch2014type}
since this helps in filtering out candidate relation types which do not 
follow this type constraint. Furthermore, for question answering 
fine-grained Named Entity Recognition (FgNER) can provide additional 
information helping to match questions to its potential answers thus improving performance~\cite{dong2015hybrid}.
For example, Li and Roth~\cite{li2002learning} rank questions based on their expected answer types 
(i.e. will the answer be \textit{food}, \textit{vehicle} or \textit{disease}).

Typically, FgNER systems use over a hundred labels,  arranged in a hierarchical structure.
We find that available training data for FgNER typically contain noisy labels, 
and creating manually annotated training data for FgNER is a
time-consuming process. Furthermore, human annotators will have 
to assign a subset of correct labels from hundreds of possible labels 
making this a somewhat arduous task. Currently, 
FgNER systems use distant supervision~\cite{craven1999constructing} to automatically generate training data.
Distant supervision is a technique which maps each entity in the corpus to knowledge bases
such as Freebase~\cite{bollacker2008freebase}, DBpedia~\cite{auer2007dbpedia}, YAGO~\cite{suchanek2007yago}
and helps with the generation of labeled data.
This method will assign the same set of labels to all mentions of a particular entity in the corpus.
For example, ``Barack Obama'' is a person, politician, lawyer, and author.
If a knowledge base has these four matching labels,
the distant supervision technique will assign all of them to every mention of ``Barack Obama''.
Therefore, the training data will also fail to distinguish
between mentions of ``Barack Obama'' in all subsequent utterances.

Ling et al.~\shortcite{ling2012fine} proposed the first system for FgNER, 
where they used 112 overlapping labels with a linear classifier perceptron for multi-label classification.
Yosef et al.~\shortcite{spaniol2012hyena} used multiple binary SVM classifiers to assign entities to a set of 505 types.
Gillick et al.~\shortcite{gillick2014context} introduced context dependent FgNER and
proposed a set of heuristics for pruning labels that might not be relevant given the local context of the entity.
Yogatama et al.~\shortcite{yogatama2015embedding} proposed an embedding based model
where user-defined features and labels were embedded into a low dimensional feature space to
facilitate information sharing among labels.

Shimaoka et al.~\shortcite{shimaoka2016attentive} proposed an attentive 
neural network model which used long short-term memory (LSTMs)
to encode the context of the entity, 
then used an attention mechanism to allow the model to focus on relevant expressions
in the entity mention's context. 
To learn entity representations, we propose a scheme which is potentially more generalizable.

\subsection{Datasets}

We evaluate our model on two publicly available datasets.
The statistics for both are shown in Table~\ref{tab:dataset-statistics}.
The details of these datasets are as follows:\\

\textbf{OntoNotes:} OntoNotes 5.0~\cite{weischedel2013ontonotes} includes texts from five different text
genres: broadcast conversation (200k), broadcast news (200k), magazine (120k), newswire
(625k), and web data (300k). This dataset is annotated with 18 categories.

\begin{figure}[htbp]
        \centering
        \includegraphics[width=\linewidth]{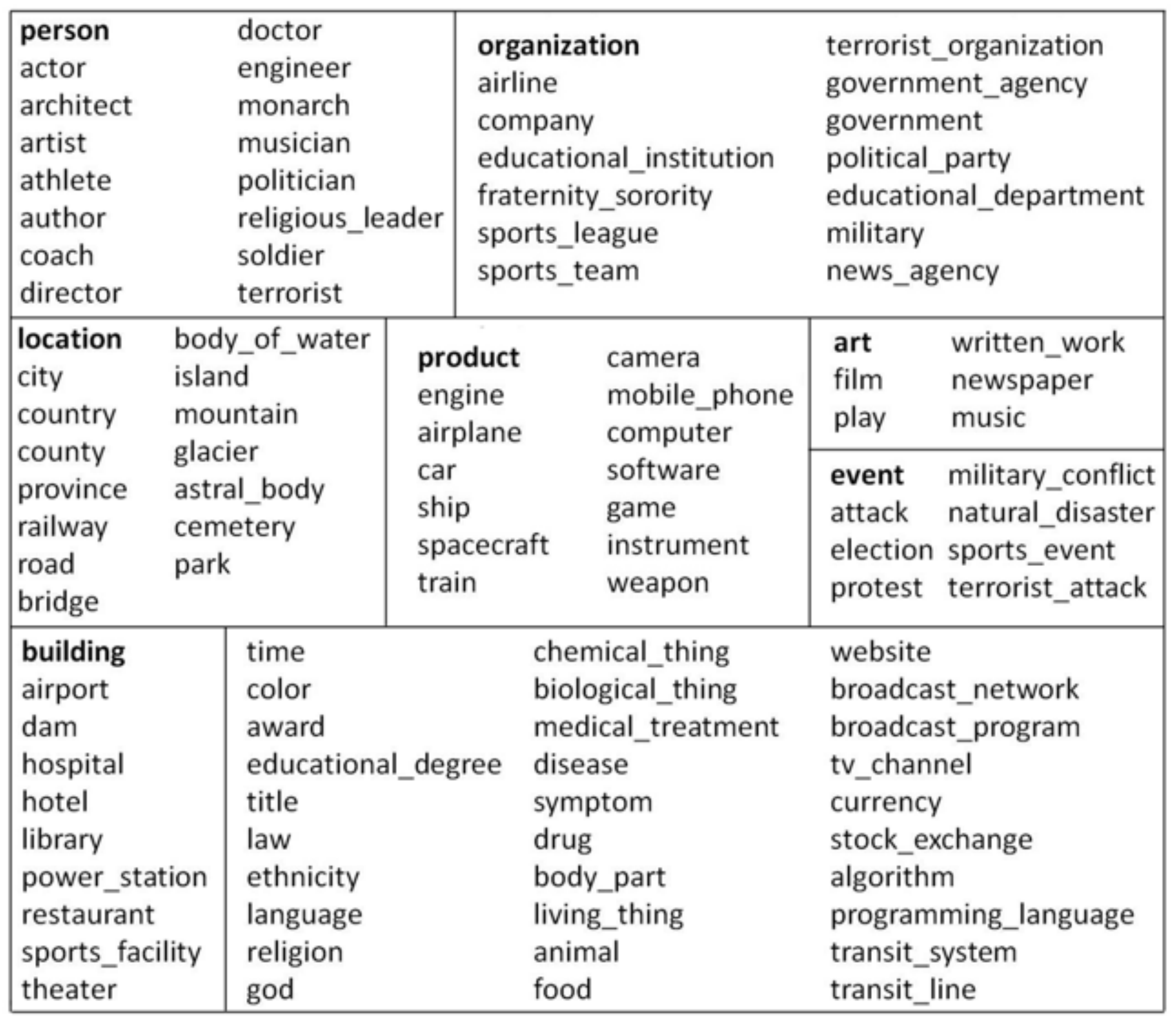}
        \caption{The 112 tags used in Wiki\textsc{(gold)}.
The tags in bold are extracted in the step described in Section 2.1. 
The finer grained tags are extracted as a final step described in Section 2.2.}
        \label{fig1}
\end{figure}

\textbf{Wiki\textsc{(gold):}} The training data consists of Wikipedia sentences
and was automatically generated using a distant supervision method,
mapping hyperlinks in Wikipedia articles to Freebase,
which we do not use in this study.
The test data, mainly consisting of sentences from news reports,
was manually annotated as described in~\cite{ling2012fine}.
The class hierarchy is shown in Figure~\ref{fig1}. 
This dataset is annotated with 7 main categories (bold text in Figure~\ref{fig1}), 
which maps directly to OntoNotes. 
The miscellaneous category in Figure~\ref{fig1} does not have direct mappings, 
so future work may include redefining these categories so the mappings are more meaningful. 

\begin{table}[]
  \centering
  \resizebox{\linewidth}{!}{%
  \begin{tabular}{|l|ll|}
  \hline
  Datasets                      & OntoNotes   & Wiki\textsc{(gold)}      \\ \hline
  \# types                      & 18          & 112     		 \\ 
  \# training labels            & 239,617     & NA 	    		 \\
  \# evaluation labels          & 23,325      & 5,943       		 \\ \hline
\end{tabular}%
}
\caption{Statistics of the datasets used in this work.}
\label{tab:dataset-statistics}
\end{table}

\subsection{Evaluation Metrics}

NER involves identifying both entity boundaries and entity types.
With ``exact-match evaluation'', a named entity is considered
correctly recognized only if both the boundaries and type
match the ground truth~\cite{ling2012fine,yogatama2015embedding,shimaoka2016attentive}.
Precision, Recall, and F-1 scores are computed on the number of
true positives (TP), false positives (FP), and false negatives (FN). Their formal definitions are as follows:

\begin{itemize}
\item True Positive (TP): entities that are recognized by NER and match the ground truth.
\item False Positive (FP): entities that are recognized by NER but do not match the ground truth.
\item False Negative (FN):  entities annotated in the ground which that are not recognized by NER.
\end{itemize}

Precision measures the ability of a NER system to present only correct entities, 
and Recall measures the ability of a NER system to recognize all entities in a corpus.
\begin{equation}
\text{Precision} = \frac{TP}{TP+FP}  \quad\quad   \text{Recall} =  \frac{TP}{TP+FN} \notag
\end{equation}

The F-1 score is the harmonic mean of precision and recall, 
and the balanced F-1 score is the variant which is most commonly used. This is defined as:
\begin{equation}
\text{F-1 score} = 2 \times \frac{\text{Precision}\times \text{Recall}}{\text{Precision}+ \text{Recall}} \notag
\end{equation}

Since most NER systems involve multiple entity types,
it is often required to assess the performance across all entity classes.
Two measures are commonly used for this purpose: the macro-averaged F-1 score and the micro-averaged F-1 score.
The macro-averaged F-1 score computes the F-1 score independently for each entity type,
then takes the average (hence treating all entity types equally).
The micro-averaged F-1 score  aggregates the contributions of entities
from all classes to compute the average (treating all entities equally).
We use the micro-averaged F-1 in our study since this accounts for label 
imbalances in the evaluation data and therefore a more meaningful statistic.

\section{Method}

Over the few past years, the emergence of deep neural networks has
fundamentally changed the design of entity detection systems.
Consequently, recurrent neural
networks (RNN) have found popularity in the field since they are able to learn long term
dependencies of sequential data.
The recent success of neural network based architectures
principally comes from its deep structure.
Training a deep neural network, however, is a difficult problem
due to vanishing or exploding gradients.
In order to solve this,  LSTMs were proposed.
An LSTM is an internal memory cell controlled by forget gate
and input gate networks. A forget gate in an LSTM layer which 
determines how much prior memory should be passed into
the next time increment. Similarly, an input gate scales new input to
memory cells. Depending on the states of both gates, LSTM
is able to capture long-term or short-term dependencies for sequential data. This is an ideal property for many NLP tasks.

\subsection{NER using ELMo}

Recently, Peters et al.~\cite{peters2018deep} proposed ELMo word representations.
ELMo extends a traditional word embedding model with features produced
bidirectionally with character convolutions.
It has been shown that the utilization of ELMo for different
NLP tasks result in improved performance compared to other types of word embedding models
such as Word2Vec~\cite{word2vec}, GloVe~\cite{ma2016end}, and fastText~\cite{wang2018code}.

\begin{figure*}[htbp]
        \centering
        \includegraphics[width=\linewidth]{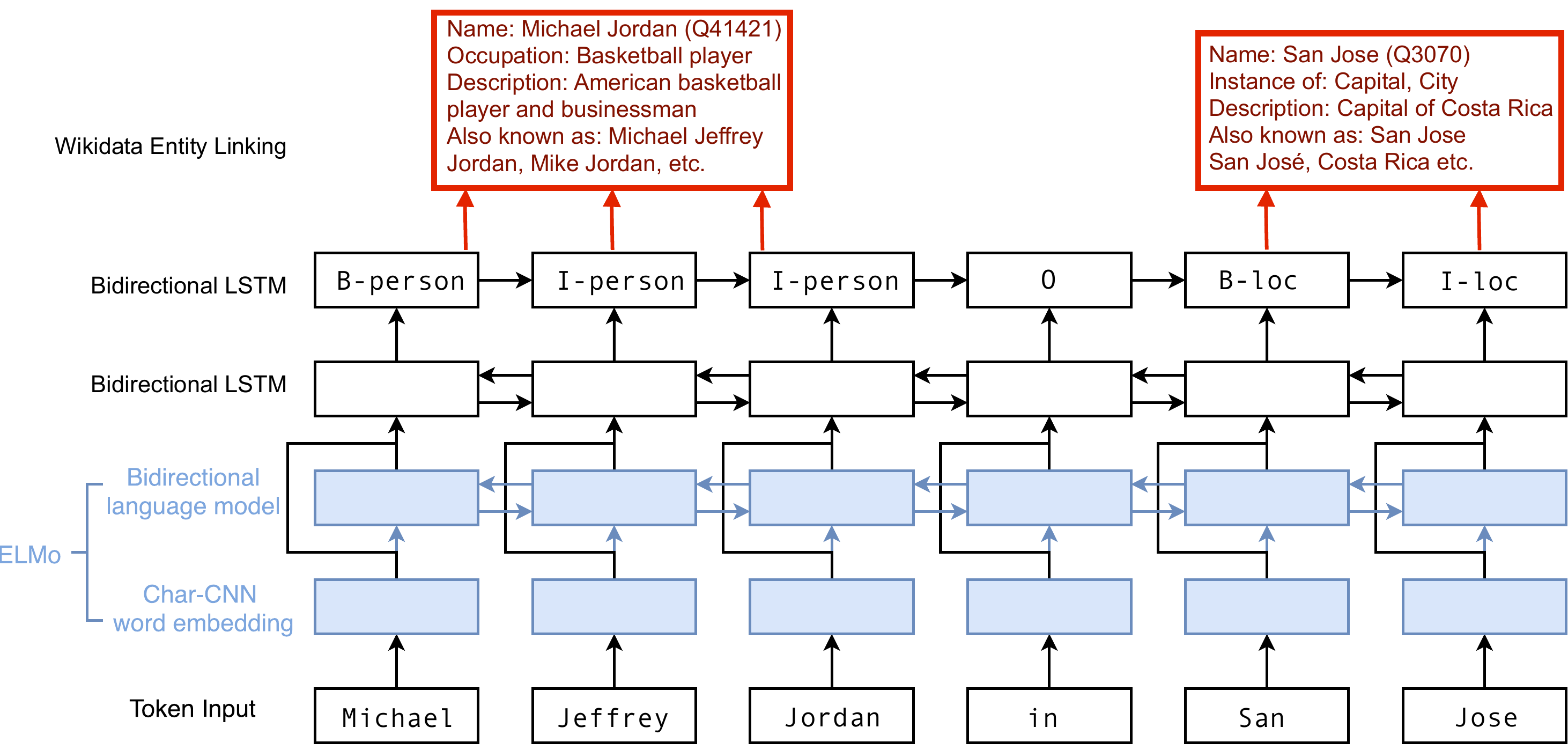}
        \caption{The full model pipeline. The first level involves token embeddings from 
                 ELMo which are fed into a residual LSTM module. 
                 The final layer involves passing the detected entities into a knowledge base, 
                 which in our case is Wikidata.}
        \label{fig2}
\end{figure*}

The architecture of our proposed model is shown in Figure~\ref{fig2}.
The input is a list of tokens and the output are the predicted entity types.
The ELMo embeddings are then used with a residual LSTM to learn informative morphological 
representations from the character sequence of each token. 
We then pass this to a softmax layer as a tag decoder to predict the entity types.

\textbf{Hyperparameter settings:}
The hidden-layer size of each LSTM within the model is set 512.
We use a dropout with the probability of 0.2 on the output of the LSTM encoders.
The embedding dimension from ELMo is 1024. The optimization method we use is Adam~\cite{kingma2014adam}.
We train with a batch size of 32 for 30 epochs.
The model was implemented using the TensorFlow\footnote{\url{http://tensorflow.org/}} framework.

\subsection{Entity Linking using Wikidata}

Entity linking (EL)~\cite{shen2018shine}, also known as named entity disambiguation or normalization,
is the task to determine the identity of entities mentioned in a piece of text with reference to a knowledge base.
There are a number of knowledge bases that provide a background repository for entity classification of this type.
For this study, we use Wikidata, which can be seen diagrammatically in Figure~\ref{fig2}. 
Systems such as DeepType~\cite{deeptype} integrate symbolic information into the reasoning process of a 
neural network with a type system and show state-of-the-art performances for EL. 
They do not, however, quote results on Wiki\textsc{(gold)} so a direct comparison is difficult. 

While these knowledge bases provide semantically rich and fine-granular classes and relationship types,
the task of entity classification often requires associating coarse-grained classes with discovered surface forms of entities.
Most existing studies consider NER and entity linking as two separate tasks, whereas we try to combine the two.
It has been shown that one can significantly increase the semantic information carried by a 
NER system when we successfully linking entities from a deep learning method to the related entities from a 
knowledge base~\cite{ji2016joint,phan2018pair}.

\textbf{Redirection:}
For the Wikidata linking element, we recognize that the lookup will be constrained by the most common lookup name for each entity.
Consider the utterance (referring to the NBA basketball player) from Figure~\ref{fig2} ``Michael Jeffrey Jordan in San Jose'' as an example. The lookup for this entity in Wikidata is ``Michael Jordan'' and consequently will not be picked up if we were to use an exact string match.
A simple method to circumvent such a problem is the usage of a
redirection list. Such a list is provided on an entity by entity basis in the ``Also known as'' section in Wikidata.
Using this redirection list, when we do not find an exact string match improves the recall of our model by 5-10\%.
Moreover, with the example of Michael Jordan (\textit{person}), using our current framework, 
we will always refer to the retired basketball player (Q41421). 
We will never, for instance, pick up Michael Jordan (Q27069141) the American football cornerback. 
Or in fact any other Michael Jordan, famous or otherwise.
One possible method to overcome this is to add a disambiguation layer,
which seeks to use context from earlier parts of the text. 
This is, however, work for future improvement and we only consider the most common version of that entity.

\textbf{Clustering:}
The Wikidata taxonomy provides thousands of possible \textit{instance of}, and \textit{subclass of} types for our entities.
Consequently, in order to perform a meaningful validation of our model, 
we must find a way to cluster these onto the 112 types provided by Wiki\textsc{(gold)}. 
Our clustering is performed as follows:

\begin{enumerate}
\item If the entity type is either \textit{person}, \textit{location}, \textit{organization}
we use the NECKAr~\cite{neckar} tool to narrow down our list of searchable entities.
\item We then look at either the \textit{occupation} for \textit{person},
or \textit{instance of} for \textit{location}/\textit{organization} categories to map to the available subtypes.
\item If the entity type is not \textit{person}, \textit{location}, or \textit{organization}
we search all of Wikidata. 
\item The clustering we perform in part 1 or 2 is from a cosine similarity of the entity description to the
list of possible subtypes for that entity. For this we use Word2Vec word embeddings trained on Wikipedia. 
We set the minimum threshold of the average cosine similarity to be 0.1.
\end{enumerate}

As an example, consider the
test sentence: ``The device will be available on sale on 20th April 2011 on amazon uk Apple's iPad''
from Figure~\ref{fig3}.
First, we tag iPad as \textit{product} using the context encoder described in Section 2.1.
We then search Wikidata and return the most common variant 
for that entity in this case Q2796 (the most referenced variant is the one with the lowest Q-id). 
We then calculate a cosine similarity of the description, in this case ``line of tablet computers'', 
with the possible subtypes of \textit{product}.
The possible subtypes, in this case, are \textit{engine, airplane, car, ship, spacecraft, train, camera, 
mobile phone, computer, software, game, instrument, ship, weapon}.
We return the highest result above 0.1, which in this case is \textit{computer} (0.54).

\section{Results}

\begin{figure}[htbp]
        \centering
        \includegraphics[width=\linewidth]{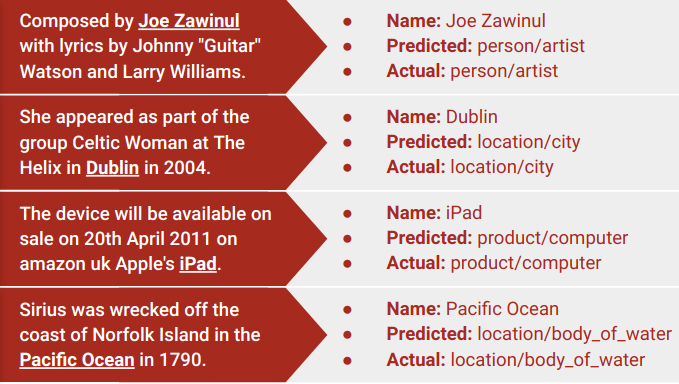}
        \caption{Some example outputs from the full model 
                 pipeline on the Wiki\textsc{(gold)} evaluation set.}
        \label{fig3}
\end{figure}

The results for each class type are shown in Table~\ref{tab2}, 
with some specific examples shown in Figure~\ref{fig3}.
For the Wiki\textsc{(gold)} we quote the micro-averaged F-1 scores for the entire top level entity category.
The total F-1 score on the OntoNotes dataset is 88\%, and the total F-1 cross-validation 
score on the 112 class Wiki\textsc{(gold)} dataset is 53\%.
It is worth noting that one could improve Wiki\textsc{(gold)} results by training directly using this dataset.
However, the aim is not to tune our model specifically on this class hierarchy. 
We instead aim to present a framework which can be modified easily to any domain 
hierarchy and has acceptable out-of-the-box performances to any fine-grained dataset.
The results in Table~\ref{tab2} (OntoNotes) only show the main 7 
categories in OntoNotes which map to Wiki\textsc{(gold)} for clarity. 
The other categories (\textit{date, time, norp, language, ordinal, cardinal, quantity, percent, money, law}) 
have F-1 scores between 80-90\%, with the exception of \textit{time} (65\%)

\begin{table}[h]
  \centering
  \resizebox{\columnwidth}{!}{
  \begin{tabular}{|l|cc|cccc|}
  \hline
  \multirow{2}{*}{Label}      & \multicolumn{2}{c|}{\bf OntoNotes} 	 &\multicolumn{4}{c|}{\bf Wiki\textsc{(gold)}} \\ \cline{2-7} 
                                    &  {\bf \%} & {\bf F-1}     & {\bf\%}   & {\bf Prec.}	& {\bf Rec.}  & {\bf F-1}   \\ \hline
  \textbf{person}                   &  14  	& 90          	& 23 & 79 	& 59 	& 66  \\ \hline
  \textbf{location}                 &  14  	& 93   		& 37 & 62 	& 47 	& 54  \\ \hline
  \textbf{organization}             &  24  	& 85  		& 26 & 45 	& 16 	& 23  \\ \hline
  \textbf{event}                    &  1   	& 70   		& 2 & 81 	& 17 	& 28   \\ \hline
  \textbf{product}                  &  1   	& 56   		& 2 & 44 	& 4 	& 8   \\ \hline
  \textbf{building}                 &  1   	& 65   		& 4 & 81 	& 17 	& 11  \\ \hline
  \textbf{art}                      &  2   	& 54   		& 0 & 0 	& 0 	& 0   \\ \hline
  \end{tabular}
}
\caption{Performance of our model from the NER classifier evaluated on OntoNotes, and the 112 subclass Wikidata linking step evaluated on Wiki\textsc{(gold)}.
         The first column denotes the percentage breakdown per class type.
         The precision, recall, and F-1 scores are shown for Wiki\textsc{(gold)}.
         For OntoNotes the precision and recall are identical for each category, therefore we only quote F-1. All values are quoted as a percentage and rounded to the nearest whole number. Since the table only shows 7 categories, 
the percentages will not sum to 100.}

\label{tab2}
\end{table}

\begin{table}[]
  \centering
  \resizebox{\linewidth}{!}{%
  \begin{tabular}{|l|l|l|}
  \hline
  Datasets                      & OntoNotes   & Wiki\textsc{(gold)}      \\ \hline
  Our model                     & 88.7\%      & 52.8\%     		 \\ 
  Akbik et al. (2018)     	& 89.7\%      & NA 	    		 \\
  Link et al. (2012)  	        & NA          & 53.2\%       		 \\ \hline
\end{tabular}%
}
\caption{Comparison with existing models.}
\label{tab3}
\end{table}

\section{Conclusion and Future Work}

In this paper, we present a deep neural network model for the task of fine-grained 
named entity classification using ELMo embeddings and Wikidata.
The proposed model learns representations for entity mentions based on
its context and incorporates the rich structure of 
Wikidata to augment these labels into finer-grained subtypes. 
We can see comparisons of our model made on Wiki\textsc{(gold)} in Table~\ref{tab3}.
We note that the model performs similarly to existing systems 
without being trained or tuned on that particular dataset.
Future work may include refining the clustering method described in Section 2.2 to extend to types
other than \textit{person}, \textit{location}, \textit{organization}, and also to include 
disambiguation of entity types.

\bibliography{eacl2017}
\bibliographystyle{eacl2017}

\end{document}